\documentclass[12pt]{iopart}
\usepackage{citesort}
\usepackage{graphicx}
\usepackage{braket}

\usepackage{iopams}

\begin{document}

\title{Universal gate-set for trapped-ion qubits using a narrow linewidth diode laser}

\author{Nitzan Akerman, Nir Navon\footnote{Present address: Cavendish Laboratory, University of Cam-
		bridge, J.J. Thomson Ave., Cambridge CB3 0HE, United Kingdom.},
	   Shlomi Kotler \footnote{Present address: Physical Measurement Laboratory, National
	   	Institute of Standards and Technology, Boulder, Colorado 80305,
	   	USA.}, Yinnon Glickman and Roee Ozeri}
\address{Department of Physics of Complex Systems, Weizmann Institute of Science, Rehovot 76100, Israel}

\ead{Nitzan.akermna@weizmann.ac.il}
\begin{abstract}
We report on the implementation of a high fidelity universal gate-set on optical qubits based on trapped $^{88}$Sr$^+$ ions for the purpose of quantum information processing. All coherent operations were performed using  a narrow linewidth diode laser. We employed a master-slave configuration for the laser, where an ultra low expansion glass (ULE) Fabry-Perot cavity is used as a stable reference as well as a spectral filter. We characterized the laser spectrum using the ions with a modified Ramsey sequence which eliminated the affect of the magnetic field noise. We demonstrated high fidelity single qubit gates with individual addressing, based on inhomogeneous micromotion, on a two-ion chain as well as the M{\o}lmer-S{\o}rensen two-qubit entangling gate.
\end{abstract}

\maketitle

\section{Introduction}
The coherent manipulation of systems at their quantum level is an essential  capability in many fields such as quantum information processing (QIP), quantum simulation and metrology. One aspect that separates a quantum system from its classical counterpart, is an exponential increase of state space as a function of the physical size of the system. The realization of arbitrary quantum operations on large systems is therefore a complicated task. Nevertheless, as in classical computation, any complex quantum operation can be approximately performed, with arbitrary accuracy, by concatenating operations taken from a finite set of simple gates. The finite set of gates from which any unitary operator can be constructed is called a universal gate set \cite{Bareno1995Elementary,zhang2003exact,vidal2004universal}. In the case of N two-level systems (a register of N-qubits), a universal gates set, for instance, may be comprised of a discrete and finite set of single qubit rotations and a two-qubit controlled-not (CNOT) gate \cite{home2009complete}. 
While it should be possible to apply each of these gates to any qubit, or qubit-pair, in the register, the number of gate types does not depend on the number of qubits involved.

Trapped-ion qubits are one of the most promising platforms for QIP. Ions can be trapped for long times and laser-cooled  to their ground motional state. Their internal electronic state as well as their motion can be controlled with high precision using lasers, microwave and radio-frequency fields \cite{haffner2008quantum,ospelkaus2011microwave}. Owing to the  isolation of trapped ions from their surrounding, quantum states can have an extremely long coherence time \cite{langer2005long}. A convenient choice of internal states for information encoding are the hyperfine and Zeeman states in the electronic ground manifold. These states are resilient against spontaneous decay and benefit from, commercially available, high-quality frequency sources. However, addressing individual ions and producing spin dependent forces presents great challenges due to the long wavelength involved. The common solution to this problem involves the use of Raman transitions, which implies relatively high intensity laser light. Another, recently proposed and demonstrated solution, is the use of near field microwave gradients \cite{warring2013individual}. A different choice of qubits are optical qubits, in which the two qubit levels are separated by an optical transition. Several of the heavier ions, that are used for QIP purposes, have low lying D levels (lower in energy than the P manifold) with a life-time of the order of one second due to the forbidden dipole transition to the S ground state. Here, the qubit levels are  encoded in the S and D manifolds which can be coupled by a narrow optical electric-quadrupole transition. The transfer of the transition photon momentum to optical qubits is straightforward and so is individual ion-qubit addressing with spatial resolution that is limited by the optical radiation wavelength. These advantages are compensated by the fact that the coherence time of such qubits is limited by the linewidth of the laser that serves as a local oscillator.  

\begin{figure}[t]
\centering
\includegraphics[width=12 cm]{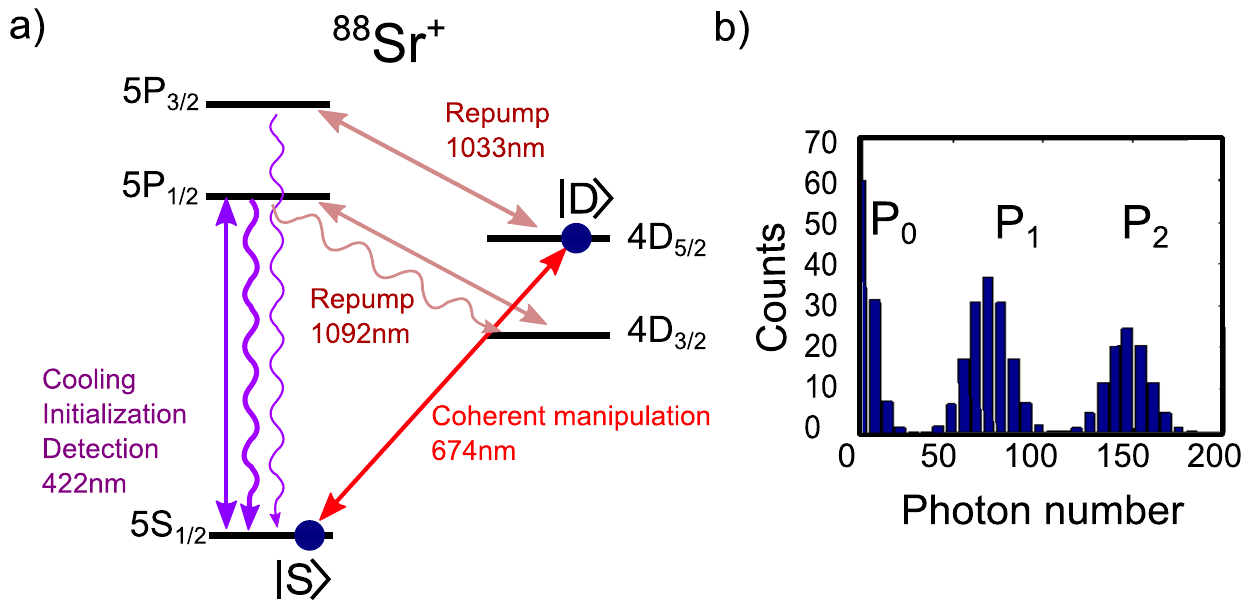}\\ 
\caption{ a) Energy level diagram and relevant transitions in $^{88}$Sr$^+$ ion. Straight arrows indicate transitions that are driven by lasers, wiggled arrows indicate transitions due to spontaneous emission, where the thick wiggled arrows represents fluorescence which is used for detection. b) An example of a measured two-qubit fluorescence histogram.}
\label{SrontiumScheme}
\end{figure}

In this paper we present a complete set of operations on a two-qubit register realized of two trapped-ion optical qubits. These includes single qubit rotations and a two-qubit entangling gate. Our main focus here is on the quadrupole transition at 674 nm. A description of the ion trap and laser systems is detailed  in \cite{akerman2012quantum}.
The relevant internal states and transitions in a $^{88}$Sr$^+$ ion are depicted in Fig.\ref{SrontiumScheme}. We encoded quantum information in two optically separated manifolds in the $\ket{S}=\ket{S_{1/2,m=+1/2}}$ and $\ket{D}=\ket{D_{5/2,m=+3/2}}$ states. The two manifolds are coupled by a quadrupole transition where the excited $D_{5/2}$ level has a lifetime of 390 ms. Driving coherent operations between the two manifolds was accomplished using a narrow linewidth 674 nm laser in a master-slave configuration. Due to the finite Lamb-Dick parameter $\eta_{674}\approx0.04$ the motional degrees of freedoms of the ion can be accessed as well. This is a necessary capability since getting the two qubits to interact can be efficiently done only through their Coulomb interaction. The scheme that was implemented here for entangling two ions, is the M{\o}lmer-S{\o}rensen (MS) gate. Incoherent operations such as cooling, preparation and detection were performed on the  $S_{1/2} \rightarrow P_{1/2}$ electric-dipole transition using a 422 nm light in addition to two repumps at 1092 nm and 1033 nm in order to deplete the population in the $D_{5/2}$ and $D_{3/2}$ levels. State initialization to the $\ket{S}$ state was done by optical pumping with circularly polarized light. Detection was carried out by counting fluorescence photon using  a photomultiplier tube for a typical duration of 1 ms. State discrimination relies on the fact that only population in the $S_{1/2}$ contributes to the fluorescence. From the statistics of the counted photons the probabilities of having zero, one and two qubits (for two-qubit register) in the $\ket{S}$ state were inferred. An example of such histogram is shown in Fig.\ref{SrontiumScheme}b. 

Manipulating optical qubits requires a narrow linewidth laser. As an example, the dephasing time of optical qubits will be limited by the inverse of the laser linewidth as the laser itself constitutes the local oscillator reference. Stabilized diode lasers are a relatively simple and cheap option for the realization of narrow linewidth lasers. Their use has been growing in popularity in the field of precision measurements and in particular, in atomic clocks where sub-Hz linewidth has been obtained \cite{Zhao20104696,ludlow2007compact,alnis2008subhertz}. 
Using narrow linewidth lasers for coherent operations on trapped ions posses slightly different requirements than atomic clocks as the former is highly sensitive to the spectral purity of the laser at a larger bandwidth i.e. in some operations, as two qubit gate, the laser is tuned to the motional sidebands which are few MHz away from the carrier transition resonance. Therefore any spectral noise at this range will drive undesired carrier transitions that will reduce the fidelity of the gate.
In our case, the main contribution to the noise at this spectral range arises from the servo loop which locks the master laser to a stable high-finesse cavity \cite{sterr2009ultrastable}. When the fast feedback is based on current modulation of the diode laser the typical unity gain, at which the induced noise is maximal, is at about 1-2 MHz. As we will show, a way to overcome this problem is to utilize the light that is transmitted trough the cavity. With a linewidth of several kHz, the cavity functions as a narrow optical filter which reduces the high frequency noise significantly.

\section{Narrow linewidth diode laser}

\subsection{The master laser}

\begin{figure}[t]
\centering
\includegraphics[width=14 cm]{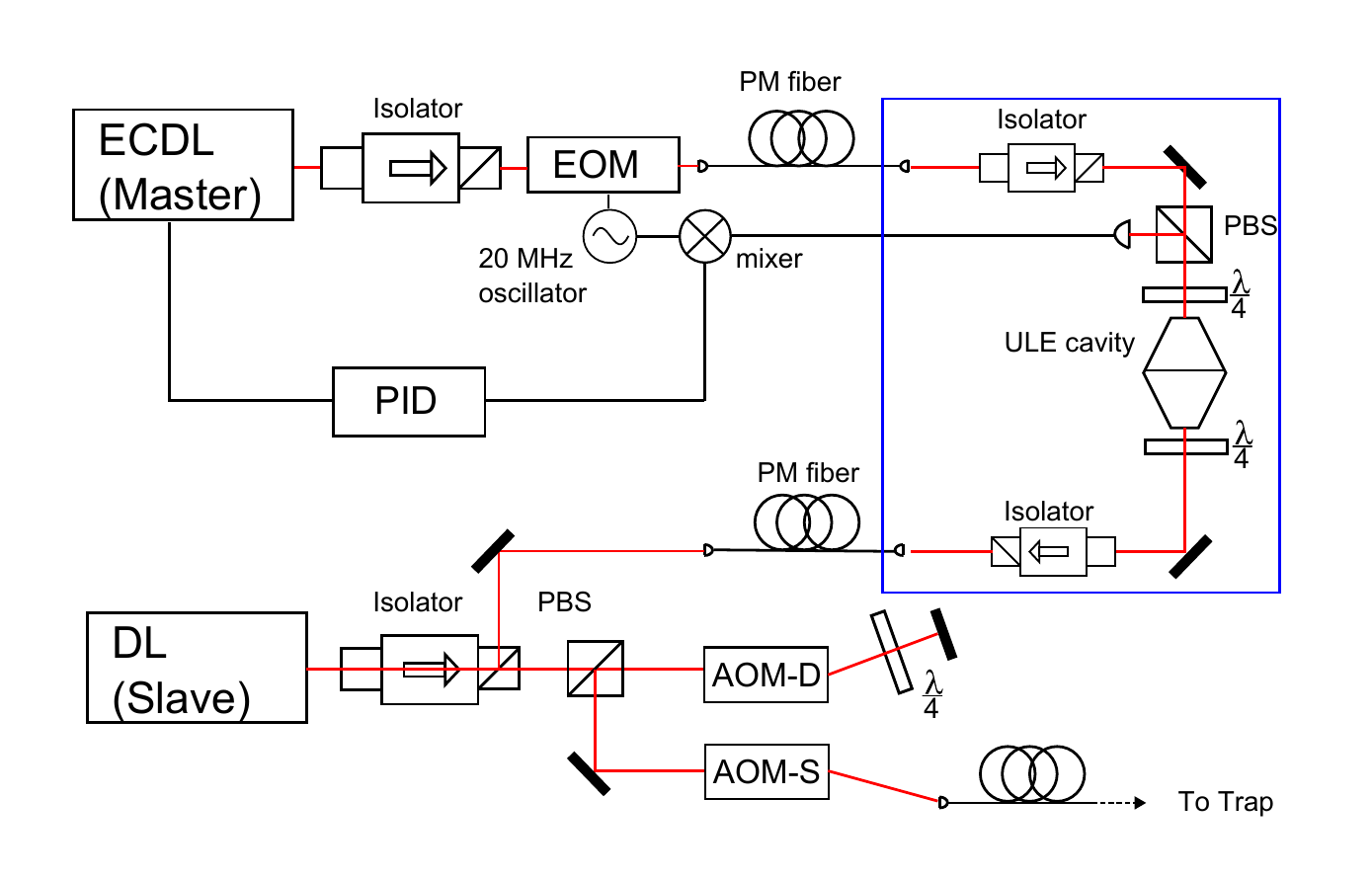}\\ 
\caption{Schematic drawing of the narrow linewidth diode-laser at 674 nm.}
\label{LaserSetup}
\end{figure}

Light is generated by an external cavity diode laser (ECDL, Toptica DL100) that is locked  to a high-finesse Fabry-Perot cavity which is made of ultra-low expansion glass (ULE). 
The ECDL, with a 10 cm long cavity, produces 10 mW of power at 674 nm and is enclosed in an acoustically-isolated box. The laser frequency is stabilized to a cavity resonance by means of the Pound-Drever-Hall (PDH) locking technique \cite{black2001introduction}. A single mode polarization maintaining (PM) fiber guides the light from within the box to the optical table where sidebands at 20 MHz are added to the laser spectrum by phase modulation using an electro-optic modulator. Roughly 500 $\mu$W of light is sent by a PM fiber to the ULE cavity which is mounted on a separate breadboard. A polarizing beam splitter (PBS) and a $\lambda/4$ retardation plate direct the reflected light from the cavity toward a 100 MHz bandwidth photo-detector to produce the PDH beat signal. The beat signal is demodulated and fed into a fast analog controller ( Toptica PDD and FALC modules) to optimize the transfer function of the servo loop. The output from the controller is fed-back into a field-effect transistor (FET) connected in parallel with the diode laser inducing current changes, which in turn steer the laser frequency. Due to the limited dynamic range of the FET, another feedback loop, implemented on a field programmable gate array (FPGA), keeps the average voltage of the FET fixed by controlling the voltage on the piezoelectric transducer that controls the ECDL grating angle.
  
The ULE reference cavity (Advanced Thin Films Inc) has a finesse of $\mathcal{F}$ = 10$^5$ measured using ring-down spectroscopy. The cavity is placed in a vacuum chamber at a pressure of $10^{-7}$ Torr. Viton rubber pads are used to reduce acoustic vibrations. The free spectral range of the cavity longitudinal modes is 1.93 GHz and the cavity linewidth is $\Gamma_c=22$ kHz. The cavity is thermally isolated by the vacuum environment and a thermal isolation material which is wrapped around the vacuum chamber. 
The resulting peak-to-peak amplitude of the cavity frequency drift at 674 nm  is about 1 MHz over a time scale of hours and the maximal slope is about 2 kHz per minute. Since the laser has to be tuned to the atomic transition frequency with sub kHz accuracy, we scan the atomic transition every three minutes and by linear interpolation evaluate the cavity resonance frequency and adjust the laser frequency accordingly at the beginning of each experiment.
\subsection{Laser frequency and amplitude control}
\begin{figure}[t!]
  \centering
  \includegraphics[width=10cm]{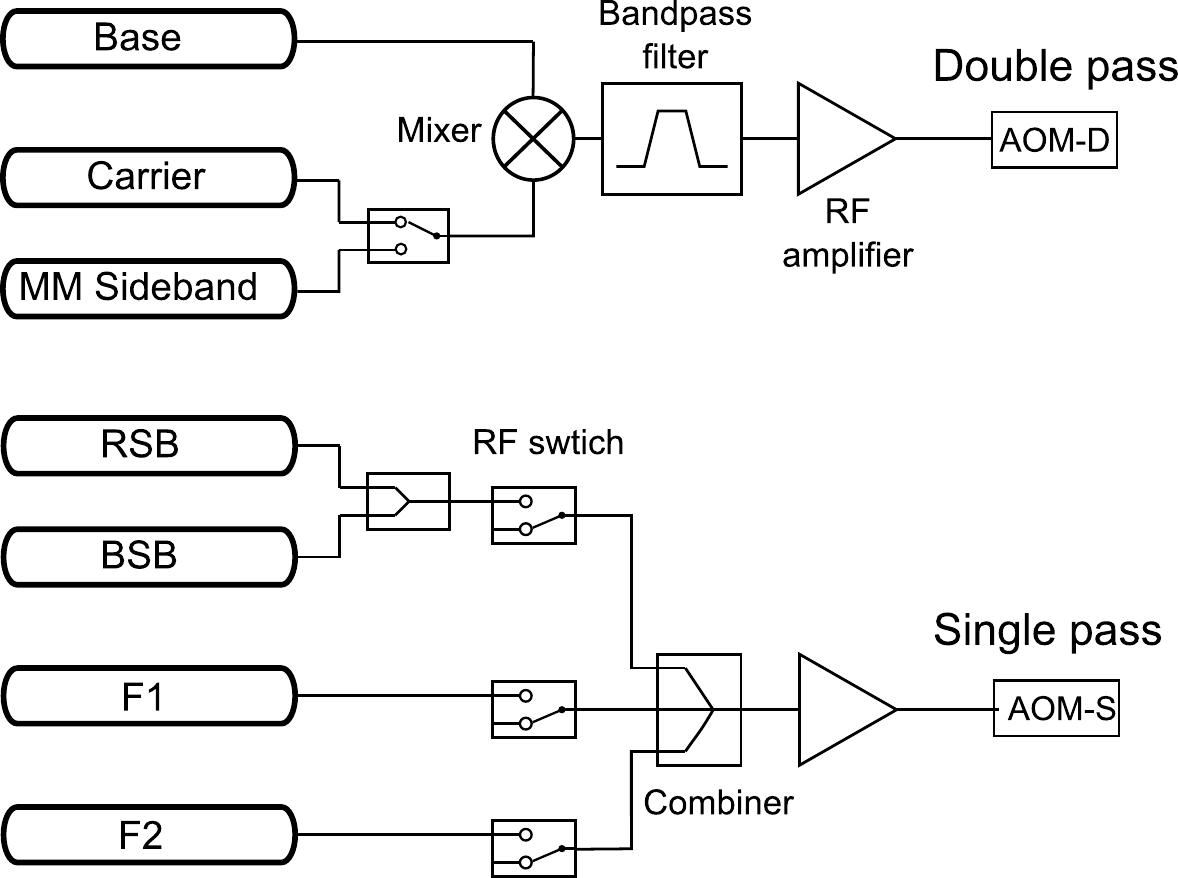}\\
  \caption{Schematic drawing of the various frequency sources for the single-pass and double-pass AOMs. The laser frequency that reaches the ions is always the sum of the double pass which switches between carrier and micromotion operations and the single pass which tunes to the carrier or vibrational sideband.}
  \label{frequencySources}
\end{figure}

The various operations on the optical qubit set a few requirements on the laser frequency control setup which is based on acousto-optic modulators (AOM):
\begin{enumerate}
\item Frequency detuning range up to the trap driving frequency of 21.75 MHZ
\item Driving two frequencies simultaneously for the MS-gate
\item Shifting all the above output frequencies to compensate for slow ULE cavity drifts 
\item Keeping a controlled  phase relation between the various frequencies 
\end{enumerate}
To satisfy all demands the use of two AOMs is unavoidable. One is in a double-pass configuration and the other is in a single-pass configuration. While the double-pass allows for a wider range of detuning it can not be used to produce a bichromatic beam. This is because when the beams are retro-reflected in the second pass they will diffract from both frequencies that drive the AOM, leading to a third beam with frequency shift of $f_1+f_2$ which is unwanted. Therefore, the bichromatic beams are generated with the single-pass AOM.
We use a separate frequency source for almost each frequency in the experiment in order to control the  phase relation between the various frequencies. The alternative of using a single frequency source and varying  its frequency  will introduce phase noise due to time jitter in the frequency updating time.
The schematic diagram of our frequency sources is presented in Fig.\ref{frequencySources}.
The double pass AOM has three frequency sources which are all based on direct digital synthesizers (DDS, Novatech 409). An rf switch is used to choose between the carrier frequency and the micro-motion (MM) sideband frequency (used for single qubit addressing which is explained in the next section) separated by 21.75/2 MHz. The two frequencies are shifted by the base frequency electronically with a mixer where the fundamental frequencies are filtered by a bandpass filter. We use the base frequency to correct for the ULE drifts without the need to update both the carrier and MM-sideband DDSs which would introduce relative phase noise between them.
The single pass AOM has four frequency sources. The red sideband (RSB) and blue sideband (BSB) generate the bichromatic beam for the two-qubit entangling gate. F1 is used to drive carrier transition with a well defined phase with respect to the RSB and BSB. The last frequency source F2 is a DDS that updates its output frequency in real-time to drive operations with no phase relation requirement such as electron shelving for state detection and ground-state cooling.  

\subsection{Laser high-frequency noise - servo bumps}

\begin{figure}[h!]
\center
\includegraphics[width=12cm]{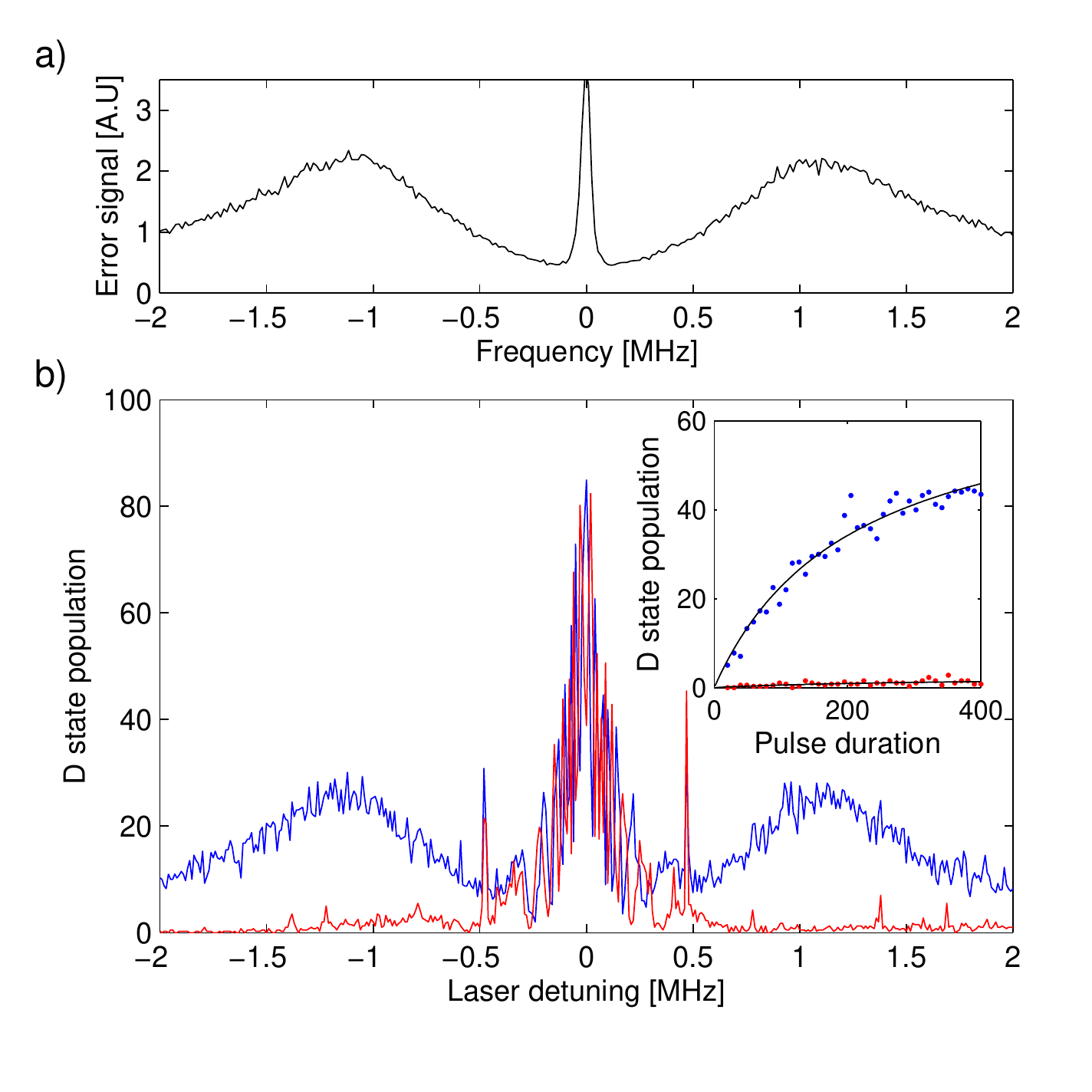}\\
\caption{Spectrum of the 674 nm laser as obtained by a) the PDH error signal and b) excitation spectroscopy on the S$\leftrightarrow$D transition with (blue) and without (red) filtering the laser with the cavity. The inset shows the excitation probability as function of the pulse duration at the peak of the servo-bump at around 1 MHz.}  
\label{MasterSpectrum}
\end{figure}

The bandwidth of the laser-cavity locking servo was found to be limited by the response of the laser diode to current modulation. Increasing the servo gain to reduce the laser linewidth is impeded by the growth of oscillations around the frequency of inverse feedback phase. Even when the gain at this frequency is kept below unity to maintain stability, noise around this frequency is amplified. These amplified noise bands around the unity gain frequency are known  also as 'servo-bumps'. In our system these servo-bumps are located at about 1-2 MHz and have FWHM of about 1 MHz.
We observed servo-bumps in the spectral content of the laser experimentally in two different ways. The first is a direct measurement of the PDH error signal spectrum with an rf spectrum analyzer. This measurement result is plotted in Fig.\ref{MasterSpectrum}a. 
Secondly, we measured the ion response to an excitation of the  quadrupole transition. Here, the laser frequency was scanned across the transition and the population in the $D_{5/2}$ state was measured. The duration of the excitation pulse was set to 100 $\mu s$, and the laser propagation direction was perpendicular to the axial direction of the trap to avoid coupling to the axial motion of the ion in the trap. The result of this measurement is presented by the blue solid line in Fig.\ref{MasterSpectrum}b. The fast oscillations around the central resonant peak are due to the 100 kHz Rabi frequency which results in many oscillations during the 100$\mu s$ pulse. The additional narrow peaks at around 0.5 MHz are due to second harmonic of a radial sideband of a transition between different Zeeman states. Both the Rabi spectroscopy and the spectrum analyzer are showing a similar feature of a 1 MHz wide response around $\pm1.1$ MHz from resonance. This wide response is attributed to incoherent excitation of the carrier transition due to the servo-bumps in laser spectrum.
This spectral feature has undesired consequences when driving vibrational sidebands transitions that are typically detuned by a 1-2 MHz from the carrier. In this case the servo-bump will drive incoherently the carrier transition and lead to decoherence. For example, we have observed that the fidelity of the MS entangling gate is severely compromised by this effect.

One way to eliminate the harmful effect of the servo-bump is to improve the performance of the servo loop by increasing its bandwidth. In this case the unity-gain frequency, and hence the servo-bumps, will be shifted towards a higher frequency and away from the motional sidebands. This can be done, for example, by introducing an intra-cavity EOM into the ECDL \cite{le2009wide}. Here, We have taken another approach in which we optically filter those servo-sidebands from the laser spectrum \cite{sterr2009ultrastable}. Fortunately, the cavity itself acts as the narrow bandpass optical filter with a 22 kHz width Lorentzian transfer-function. Hence, the light that is transmitted through the cavity possesses high spectral purity with, in principle, three orders of magnitude reduction in the noise spectral power density around the 1 MHz servo-bumps.
  
The price for relying on the transmitted light is the attenuation in the available light intensity. In our setup , the cavity transmission is only about 5\%. While a decrease by a factor of two can be accounted for spatial mode-mismatch, most of the loss is due to intra-cavity losses. In addition, the intensity of the light that is sent to the cavity is limited to the mW range in order to prevent substantial thermal effects on the cavity mirrors (ULE substrate)  that would in turn lead to thermal shifts of the cavity resonance frequency. Overall the transmitted light intensity is limited to few 10's of $\mu W$, which is well below the required intensity for ion-qubit operations, and hence needs to be amplified. To this end, we incorporate another slave diode-laser which is injection-locked to the filtered master-laser light. A slave diode rather than a tapered amplifier is chosen due to the much smaller intensity that is needed to injection lock a diode as compared with the intensity that is required to saturate a tapered amplifier.

The complete master-slave setup is shown in Fig.\ref{LaserSetup}. The transmitted light polarization is restored to linear by another $\lambda/4$  plate and coupled to a PM-fiber. Two optical isolators are placed at the input and output of the cavity to prevent optical interference. Between 5-10 $\mu W$ of filtered light is injected to the slave diode laser using  mode-matching optics. This intensity is sufficient for a stable operation of the slave laser with an injected current window of 0.1 mA.

Intensity fluctuations of the 674 nm light lead to fluctuations in the Rabi frequency, and therefore affect any coherent operation. The origin of intensity noise is mainly due to fiber phase noise that is translated to intensity noise when passing through a polarizing beam splitter (PBS) on the way to the ion. To prevent this, the intensity of the laser is monitored close to the input port of the vacuum camber and stabilized by adjusting the rf power that drives the AOM. 

Figure \ref{MasterSpectrum}b compares between the excitation spectrum of the ion when excited with the slave laser when it is injection locked to the unfiltered master laser (blue curve) and to the cavity filtered light (red curve). The significant improvement in the spectral purity is clearly observed. The inset shows a measurement of the excitation probability as function of the pulse duration at the maximum of the servo-bump in the two cases. In the unfiltered light case already at a pulse duration of 400 $\mu s$ the population is saturated close to 50\% while in the filtered light case as little as 1-2\% are observed. The ratio between the slopes of the two curves in the linear regime is about 50 which should be regarded only as a lower bound due to the small signal in the filtered case which could be a result of power broadening together with other decoherence mechanisms such as magnetic field fluctuations or leakage of the 1033 nm repump light.


\subsection{Characterization of the laser linewidth}
In quantum information processing the qubit coherence time is determined by the stability of the relative phase between the qubit and the local oscillator. In the case of an optical qubit, the local oscillator that keeps track of the qubit phase, is a narrow linewidth laser. The physical mechanisms that cause phase fluctuations to the qubit and the laser are different and uncorrelated. Therefore, reaching a long coherence time requires that the phases of both the qubit and the oscillator are independently stable.
While commercially available oscillators at radio and microwave frequencies can easily reach coherence times of many seconds (requiring fractional frequency uncertainty of $10^{-6} - 10^{-9}$), obtaining a laser with a coherence time of one second, i.e. 1 Hz linewidth, is a much more challenging task and requires fractional frequency uncertainty of $10^{-14} - 10^{-15}$. On the qubit side, the dominant source of dephasing are fluctuations of the ambient magnetic field .

One method of characterizing the spectral properties of a laser is a delayed self-heterodyne interferometer \cite{okoshi1980novel}. Here the laser light interferes with a delayed fraction of itself by using a long optical fiber to create a substantial time delay $\tau$. However, as the spectral resolution of this method is given by $1/\tau$, to measure the sub-kHz frequency fluctuations requires hundreds of kilometers of fiber length which render this method impractical. The alternative is to compare against another stable reference. Examples for such a reference include another independent laser, an ultra stable Fabry-Perot cavity or a narrow atomic transition.

\begin{figure}
\centering
\includegraphics[width=16cm]{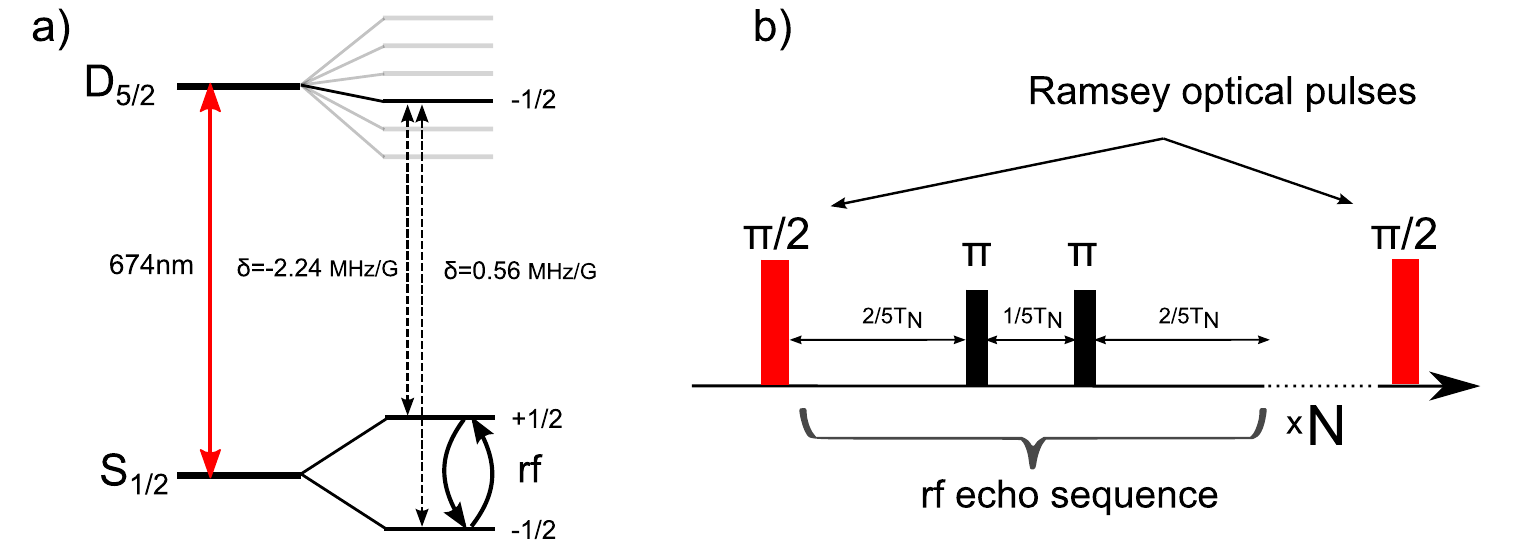}
 \caption{Magnetic field noise dynamic decoupling scheme. The Zeeman shifts of two S$_{1/2}$ ground state sub-levels with respect to the 5D$_{5/2,-1/2}$ have opposite signs. Hence, the contributions of the magnetic field noise to decoherence in a Ramsey experiment can be suppressed by applying an appropriate sequence of rf $\pi$ pulses.}
 \label{MagneticDD}
\end{figure}
Since the purpose of our laser is to drive the $S_{1/2}\rightarrow D_{5/2}$ transition in $^{88}Sr^+$ and with no other tool at hand, we inferred our  laser linewidth by performing a Ramsey spectroscopy measurements on the ion. One complication that arises in our case is that the relevant quadrupole transition is first order Zeeman sensitive so a conventional Ramsey experiment can not discriminate between the magnetic field noise and the laser frequency noise. 
Previously, custom-designed entangled state were used to differentiate laser phase noise from magnetic field noise \cite{roos2006designer} Here we demonstrate a custom-designed dynamic decoupling sequence to this end. The advantage of using dynamic decoupling for this purpose is the simplicity of its implementation. Our magnetic field dynamic decoupling (MFDD) eliminates only phase fluctuations due to slow magnetic field noise, thus rendering the optical transition to an effective clock transition.
After initializing to the superposition state $\psi=1/\sqrt{2}(\ket{S_{1/2,-1/2}}+\ket{D_{5/2,-1/2}})$ the scheme follows conventional dynamic decoupling methods \cite{viola1998dynamical} in the form of echo pulses with two main differences. The first is that the pulses are performed in the $S_{1/2}$ spin manifold and not on the optical transition as this will reverse the effect laser frequency noise as well as magnetic field noise. The basic idea is illustrated in Fig.\ref{MagneticDD} and relies on the opposite sign of the Zeeman shift of the $S_{1/2,-1/2}\leftrightarrow D_{5/2,-1/2}$ transition as compared with the  $S_{1/2,1/2}\leftrightarrow D_{5/2,-1/2}$ transition. As long as the magnetic field does not vary between echo pulses, the phase that is accumulated when the ground state population is in the $S_{1/2,-1/2}$ due to an arbitrary Zeeman detuning, can be reversed by transferring the population to the $S_{1/2,1/2}$ state where the phase is accumulated with the opposite sign. The second difference from a regular echo sequence is that the time which is spent in each state is not equal. This is because the magnitude of the Zeeman shift in the two states is different with a ratio of 1:4. 
The experimental sequence is illustrated in Fig.\ref{MagneticDD}b. For a Ramsey experiment of a total time T and N rf echo sequences, each echo sequence is of duration $T_N=T/N$ and is composed of two rf $\pi$ pulses separated by a time $T_N/5$ at the middle of the sequence.
Notice that in terms of the optical transition this scheme is identical to a conventional Ramsey sequence having only two $\pi/2$ pulse with a wait time in between where it is only the ground state part that is manipulated with rf pulses. We have shown, in previous work \cite{kotler2011single}, that the coherence of ground-state superpositions can be maintained for as long as a second using several hundred echo pulses. In general any dynamic decouple scheme can be adapted for the manipulation of the ground state part.
\begin{figure}
  \centering
  \includegraphics[width=16cm]{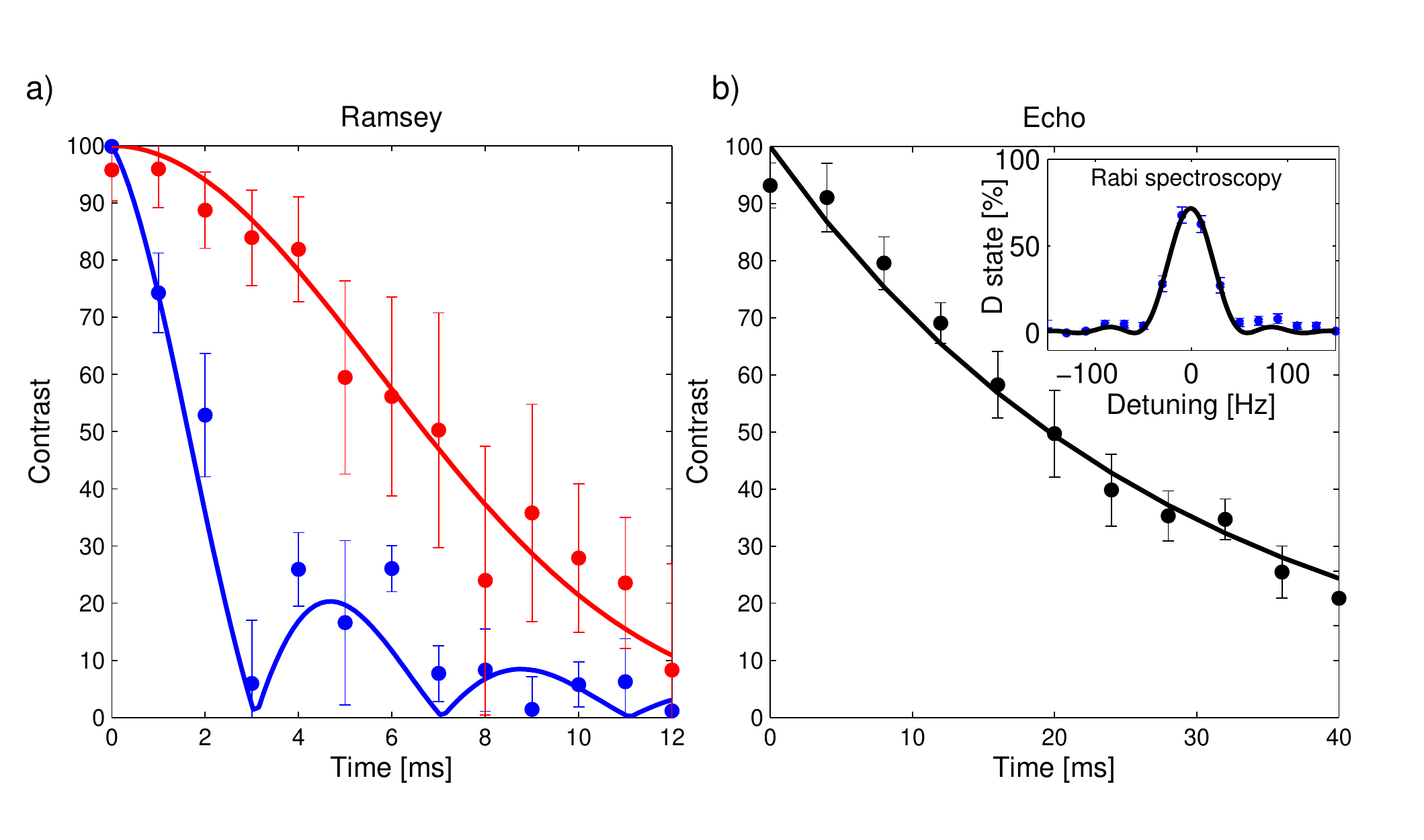}\\
  \caption{Measurement of the laser phase correlation time by a Ramsey experiment. a) Fringe contrast as function of the Ramsey time with (red) and without (blue) MFDD sequence with N=4 as shown in Fig\ref{MagneticDD}\.b. The solid blue (red) line is a theoretical
  	fit to a Bessel (Gaussian) function. b) Fringe contrast as a function of the experiment time in Hahn-echo sequence where a single echo $\pi$ pulse is performed on the optical transition. Solid black line is a theoretical fit to a decaying exponent. The Inset show a Rabi spectroscopy scan which is consistent with the result of the Ramsey and echo measurements.}
  \label{RamseyAndEcho}
\end{figure}
The experimental results are presented in Fig.\ref{RamseyAndEcho}a. We plot the Ramsey fringe contrast as a function of the total experiment time for a plain Ramsey sequence (blue dots) and MFDD sequence with N=4 (red dots). The contrast is obtained by scanning the phase of the second $\pi/2$ optical pulse and fitting the data to a sine function with the amplitude and phase as fit parameters. Solid lines are theoretical fits to a Bessel function in the plain Ramsey case and a Gaussian in the MFDD case. In the regular Ramsey experiment the contrast dropped to 50\% after 1.7 ms. Here, the 50 Hz monotone magnetic field noise component dominated and the contrast followed a Bessel shape \cite{kotler2013nonlinear}. Applying the MFDD sequence resulted in a coherence time of 6.7 ms (more than a four-fold improvement) and in addition the contrast follows a Gaussian function shape rather than a Bessel, which implies that there is no single frequency component that dominated the noise spectrum.
Applying MFDD sequences with N$>4 $ did not extend the coherence time. Hence, we infer that at this point the laser frequency noise, rather than the magnetic field noise, is the limiting factor. Assuming a Gaussian line-shape, we infer a laser linewidth of 65(5) Hz. This result is consistent with our Rabi spectroscopy measurements depicted in the inset of Fig.\ref{RamseyAndEcho}b. More information on the laser spectral content can be obtained by a standard single echo sequence on the optical transition. Here, we did not apply the MFDD sequence but rather incorporated an active stabilization system that reduced the ambient magnetic field noise to a negligible level. As shown in Fig.\ref{RamseyAndEcho}b. the contrast extended to 20 ms following exponential decay. Since the echo pulse acts as a highpass filter we infer that the coherence time measured without the echo is limited by low frequency noise which we assume to has a 1/f characteristic. If we consider the echo measurement to represent the laser fast-linewidth (white noise contribution) then from the exponential fit we obtained 11(1) Hz linewidth \cite{riehle2006frequency}.

\section{Single-qubit gate}
\begin{figure}[h!]
  \centering
  \includegraphics[width=12cm]{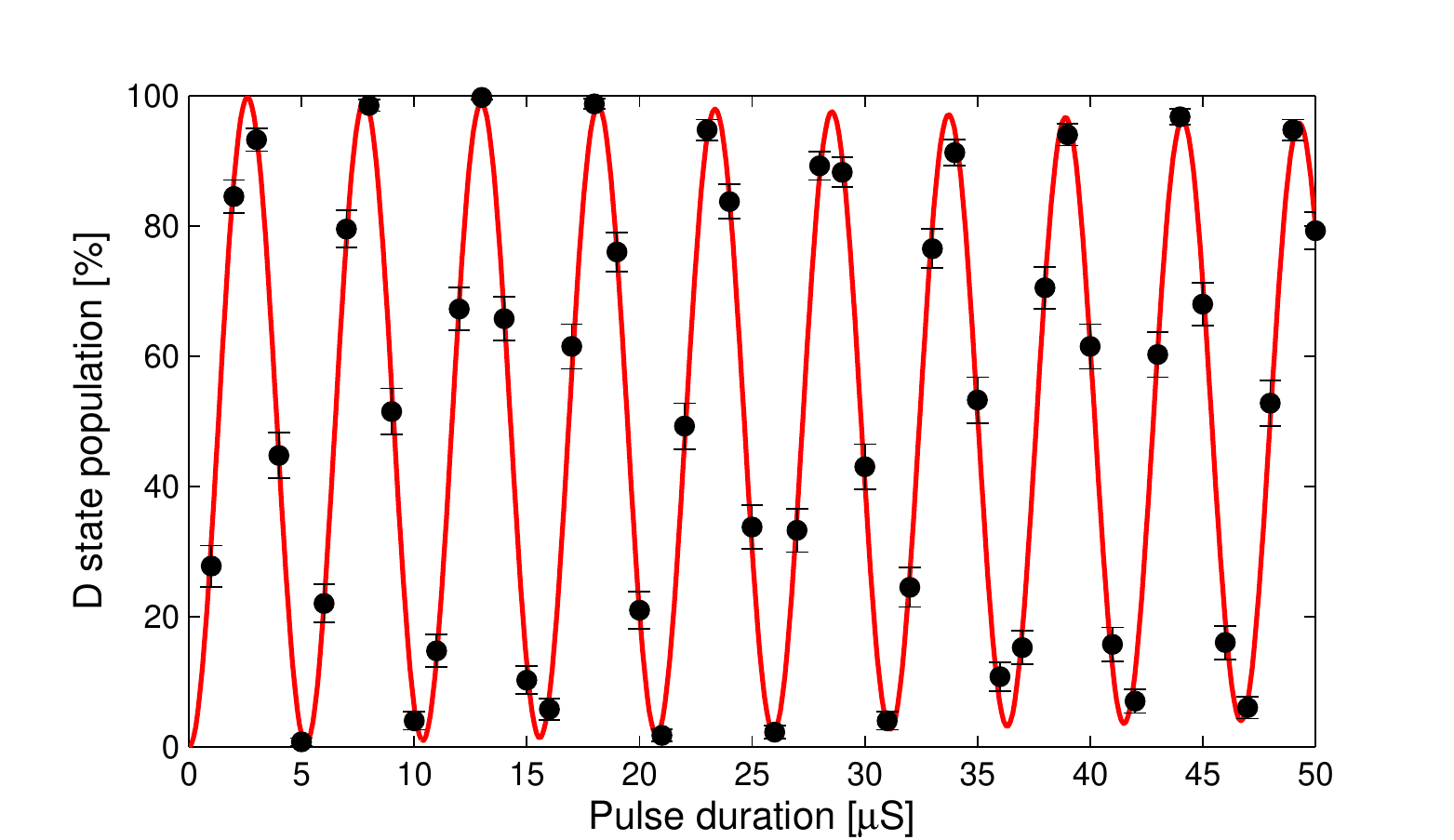}
  \caption{Collective Carrier Rabi oscillation on the $S_{1/2,1/2}\leftrightarrow D_{5/2,3/2}$ quadrupole transition in Doppler cooled ions. From the fit (solid red line) we deduce a Rabi frequency of $\Omega/2\pi=193$ kHz and a single flip fidelity of 0.997(2)}.
  \label{CarrierRabi}
\end{figure}
For the characterization of single qubit operations in the context of quantum information processing we need to distinguish between operations which are collective and act on all the qubits in the register, and  operations that are applied to individual qubits. The former is similar to the case of a single-qubit register in which high fidelity rotation gates that go beyond current estimates for fault tolerance threshold have been demonstrated \cite{Harty2014High,brown2011single}. In our case, we perform single qubit rotations by driving the optical electric quadrupole transition. Figure \ref{CarrierRabi} shows 20 consecutive collective carrier Rabi flops on a two-qubit register with Rabi frequency of $\Omega_c/2\pi \approx 200$ kHz. The small reduction in the contrast can be attributed to the laser frequency and amplitude noise as well as the motional spectator modes as the ions are only Doppler cooled with axial (radials) secular frequency of 1.95(3.8) MHz. From the fitted decaying envelope we can deduce a single flip ($\pi$ rotation) fidelity larger than 0.997(2) but a more reliable result would be inferred by performing randomized benchmarking \cite{gaebler2012randomized} which is beyond the scope of this work. 

\begin{figure}[!ht]
	\centering
	\includegraphics[width=12cm]{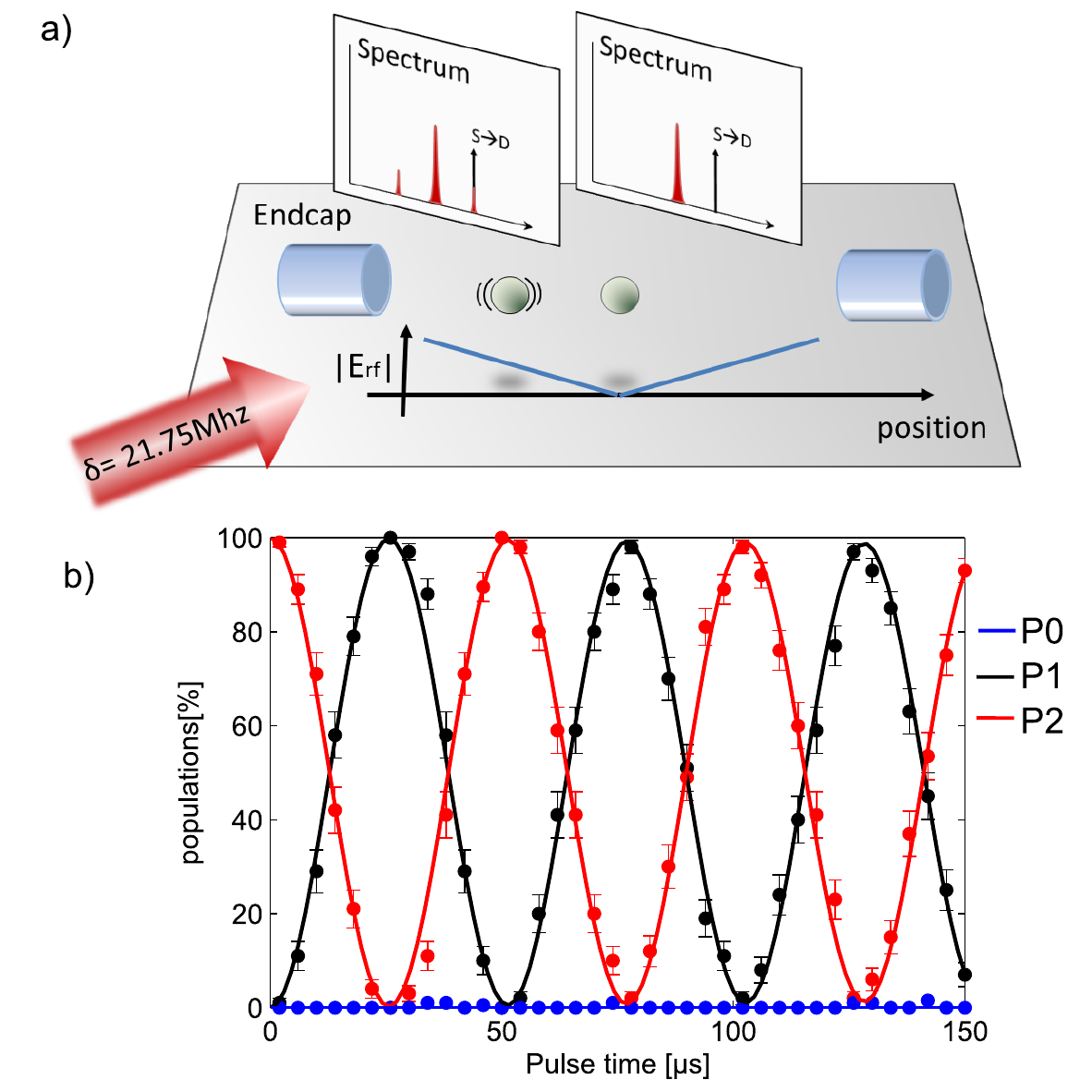}\\
	\caption{Single qubit addressing based on MM-sideband. a) Graphic representation of the single qubit addressing scheme. The two ions experience different micromotion due to their separate locations in the trap. Consequentially, the spectrum of a spatially homogeneous laser beam will be different in the ions frames of reference. Detuning the laser frequency by the trap drive frequency, results in a Rabi frequency that depends on the MM amplitude. b) One qubit Rabi oscillation in a two-ion crystal by driving the MM-sideband. The Rabi frequency is $\sim 20$ kHz.}
	\label{SingleQubitAddressing}
\end{figure}

In the case of more than one qubit, where single qubits need to be addressed individually, the experimental challenges are different. Here only a handful of experiments have demonstrated single qubit addressing and gates with high fidelities \cite{wang2009individual,piltz2014trapped,home2009complete,schindler2013quantum}. The major challenge here is to apply a high fidelity operation on one qubit while preventing unwanted state changes to the neighboring qubits. The two most popular methods rely on tightly focused laser beams that locally induce a light shift on a singe ion qubit and strong magnetic gradients, both lift the degeneracy between the different qubits in the register and allow to individually address the desired qubit in the spectral domain. Our approach uses the readily available spatial variation of the rf electric field amplitude of the ion trap along the axial direction \cite{Navon2013Addressing}. The basic principle, which is illustrated in Fig.\ref{SingleQubitAddressing}a., is to utilize the excess micromotion to drive the optical qubit transition. For the case of only two qubits, which is discussed here, we position the two-ion crystal such that one is exactly at the rf null (null-qubit) whereas the other qubit (MM-qubit), which is not at the null, undergoes micromotion at the trap drive frequency of 21.75 MHz. The excess micromotion that we use is along the axial direction of the trap and is due to the boundary conditions that are imposed by the endcap electrodes. In the ion frame of reference, the periodic Doppler shift due to excess micromotion adds sidebands to the laser spectrum. By detuning the laser frequency from the carrier transition by the trap rf frequency, the MM-sidebands can be used to  resonantly drive coherent operations on MM-qubit while the Rabi frequency for null-qubit is largely reduced and ideally nulls altogether. The Rabi frequency for driving the MM-sidebands depends on the micromotion amplitude, $\Omega_{MM}=\Omega_c\cdot J_1(\mathbf{x}_{MM}\cdot \mathbf{k})$, where $\Omega_c$ is the carrier Rabi frequency in the absence of excess micromotion, $J_1$ is the first order Bessel function, $\mathbf{k}$ is the laser wavevector and $\mathbf{x}_{MM}$ is the micromotion amplitude vector. When $\mathbf{x}_{MM}$ is much smaller than the laser wavelength the Rabi frequency can be approximated by $\Omega_{MM}=\Omega_c/2\cdot \mathbf{x}_{MM}\cdot \mathbf{k}$. Moreover, the reduction in carrier Rabi frequency due to the micromotion is second order in the amplitude and can be neglected.

Fig.\ref{SingleQubitAddressing}b shows single qubit Rabi flops by driving the  MM-sidebands. The various curves represent the different measured register states P0 (blue), P1(black) and P2(red) which correspond to the probabilities of finding zero, one and two qubits in the $\ket{S}$ state. As is expected from only one qubit flopping, P1 and P2 oscillate with an opposite phase with almost 100\% amplitude while P0 stays below 1\% throughout the measurement. This means that the null-qubit remained in the $\ket{S}$ while the MM-qubit oscillated periodically between the $\ket{S}$ and $\ket{D}$ states with a Rabi frequency of roughly 20 kHz. Here, the axial frequency was lowered to 1 MHz which increased the separation between the ions and therefore the micromotion amplitude and Rabi frequency of the MM-sidebands. In this measurement the single qubit flip ($\pi$ rotation) fidelity was as high as the collective qubits flips that are presented in Fig.\ref{CarrierRabi}. 
However, the MM-sidebands gate is not as robust and stable as collective carrier rotations. This is due to the extreme sensitivity of the excess MM to the electrostatic environment where very small changes in the ambient electric fields change the MM-sidebands Rabi frequency and introduce rotation errors. To mitigate the drift in the MM-sidebands Rabi frequency, composite pulses can be applied. As an example, the sequence $X_{\pi/2}Y_{\pi}X_{\pi/2}$ where $X,Y$ are the direction of rotation, implements a $\pi$ rotation that is first-order insensitive to small variation in the Rabi frequency.    

The two operations discussed above: two-qubit collective rotations and individual addressing of only one qubit are sufficient in the case of a two-qubit register to perform any single qubit gate. As an example in ref \cite{navon2013quantum} we have used this scheme to perform complete quantum process tomography for the M{\o}lmer-S{\o}rensen two qubit gate. In principle the use of the spatially varying electric field and micromotion induced sidebands can be generalized to more than two ions by introducing the concept of dressed-state picture \cite{Navon2013Addressing}.

\section{M{\o}lmer-S{\o}rensen two-qubit interaction}
\begin{figure}[!b]
  \centering
  \includegraphics[width=12cm]{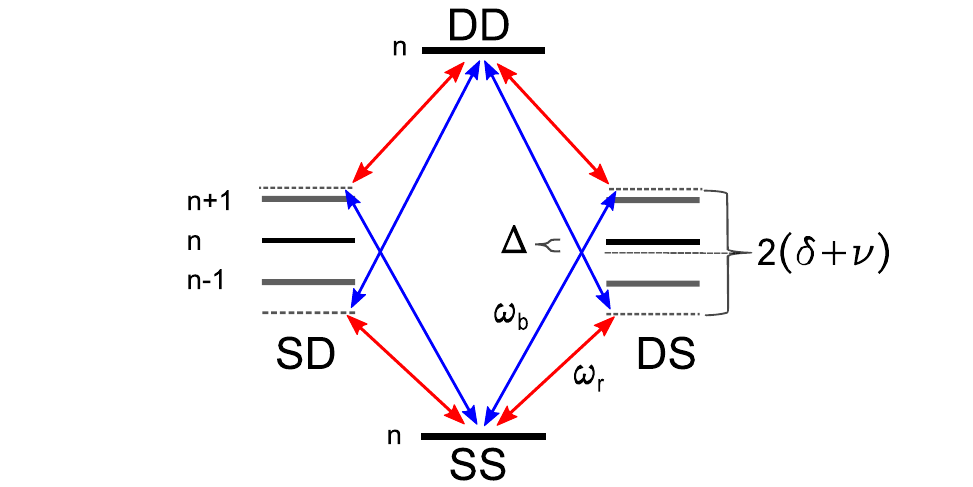}\\
  \caption{The M{\o}lmer-S{\o}rensen interaction scheme. A bichromatic beam couples the $\ket{SS,n}$ to the $\ket{DD,n}$ through the intermediate states $\ket{SD,n \pm 1}$ and $\ket{DS,n \pm 1}$. The two frequencies $\omega_{r,b}$ are ideally symmetrically detuned from the carrier transition frequency $\omega_0$ by $\delta+\nu$, where $\nu$ is the frequency of the vibrational mode through which the gate is performed. In addition the frequencies can be non-symmetrically detuned from the carrier transition by an asymmetric detuning $\Delta$.}  
  \label{MSinteraction diagram}
\end{figure}
Experimentally implementing a universal two-qubit gate is typically more challenging than implementing single-qubit rotations. This is because two-qubit gates require the synthesis of bipartite interactions between qubits. Since trapped ions are separated by a few microns, the interaction between their qubit states is synthesized by coupling their internal states to their quantized collective modes of motion. 
Here we present our implementation of the M{\o}lmer-S{\o}rensen (MS) entangling gate \cite{sorensen2000entanglement} which, together with a universal single qubit rotations, forms a universal gate-set for quantum computation. The MS scheme, besides being successful in reaching fidelities as high as 99.3\% \cite{benhelm2008towards}, holds some desirable properties such as resilience to finite temperature, straightforward generalization to multi ion-qubit gates and relative simplicity of implementation.
The MS interaction scheme is presented in Fig.\ref{MSinteraction diagram}. It consists of a bichromatic field that drives collective internal state atomic transitions through the red and blue sidebands of a single vibrational mode. While in general the MS interaction entangles the internal state of the ions with their motion, at certain periodic times the internal states and motion return to be disentangled. At those times the ions internal states undergo collective rotations and the motion returns to it initial state.    
The interaction Hamiltonian in the rotating wave approximation takes the following form \cite{roos2008ion}
\begin{equation}\label{SM-hamiltonian}
H_{int}=\frac{\hbar\Omega}{2}(i\eta\e^{-i\Delta t}(a^\dag e^{-i\delta t}+a e^{i\delta t})(\sigma^{(1)}_- +\sigma^{(2)}_-) + h.c.)
\end{equation}
where $\Omega$ is the carrier Rabi frequency and we assume an equal intensity in each of the two beams, $\eta$ is the Lamb-Dicke parameter, $\delta=(\omega_b-\omega_r)/2-\nu$ is the bichromatic field symmetric detuning from the motional sidebands, $\Delta=(\omega_b+\omega_r)/2-\omega_A$ is the bichromatic field asymmetric detuning from the two-photon transition frequency, $a$ and $a^\dag$ are the creation and annihilation operators of the vibrational normal mode with a frequency $\nu$ and $\sigma_{-,+}$ are the ladder operators for the atomic transition. 

In the absence of asymmetric detuning, $\Delta=0$, an analytic expression for the time evolution of atomic populations can be obtained \cite{kirchmair2009deterministic}. When starting from the vibrational ground state (n=0) the population dynamics takes the following form, 
\begin{eqnarray}
  P0&=&\frac{1}{8}(3+e^{2|\alpha|}-4e^{|\alpha|/2})+\sin^2(\theta)e^{-|\alpha|^2} \\
  P1&=&\frac{1}{4}(1-e^{-2|\alpha|^2}) \\
  P2&=&\frac{1}{8}(3+e^{2|\alpha|}-4e^{|\alpha|/2})+\cos^2(\theta)e^{-|\alpha|^2}
\end{eqnarray}
where 
\[ \theta(t)=\frac{\eta^2\Omega^2t}{2\delta}(1-\mbox{sinc}(\delta t)) ~~\mbox{and} ~~ \alpha(t)=\frac{\eta\Omega}{\delta}(e^{i\delta t}-1)
\]
Starting from the state $\ket{SS}$, a maximally entangled state $\ket{\Phi}=(\ket{SS}+i\ket{DD})/\sqrt2$ is obtained with a detuning of $\delta=2\eta\Omega$ and after an interaction time of $T_{gate}=2\pi/\delta$. 
In cases where the asymmetric detuning non-zero, the dynamics can be obtained by numerical integration.

\subsection{Extensive 2D scans of the MS-interaction parameters}
\begin{figure}
  \centering
  \includegraphics[width=15cm]{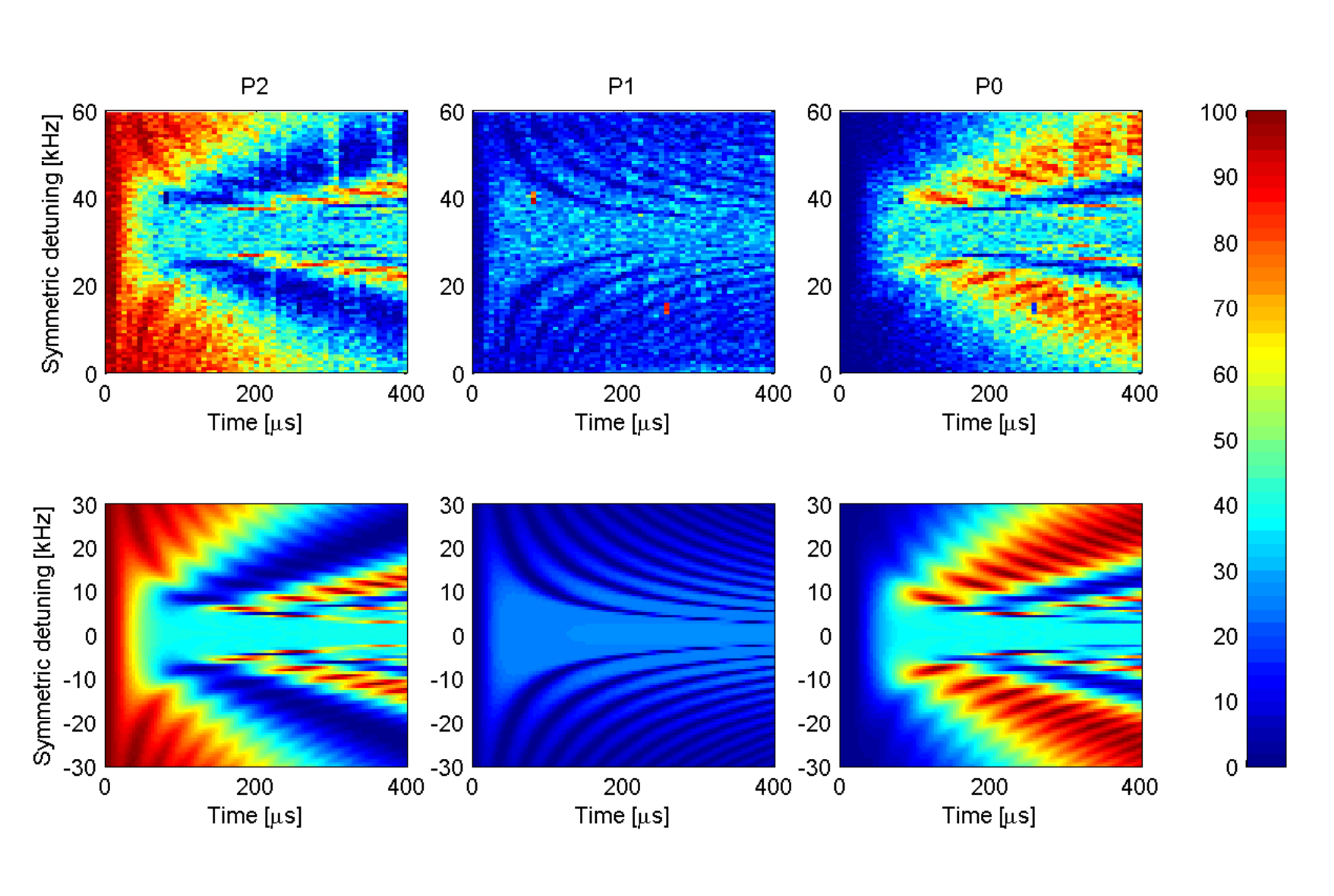}\\
  \caption{Populations time dynamic for the MS-interaction as function of the gate detuning. For constant asymmetric detuning $\Delta=0$. Upper panels are the measured populations and lower panels are the calculated result of Eqs. (2)-(4). The shift in the y axis between the measured and calculated result is exactly the uncompensated light shift contribution in the measurement of the vibrational mode which was obtained by sideband spectroscopy.}
  \label{GateMapSymetric}
\end{figure}
\begin{figure}
  \center
  \includegraphics[width=15cm]{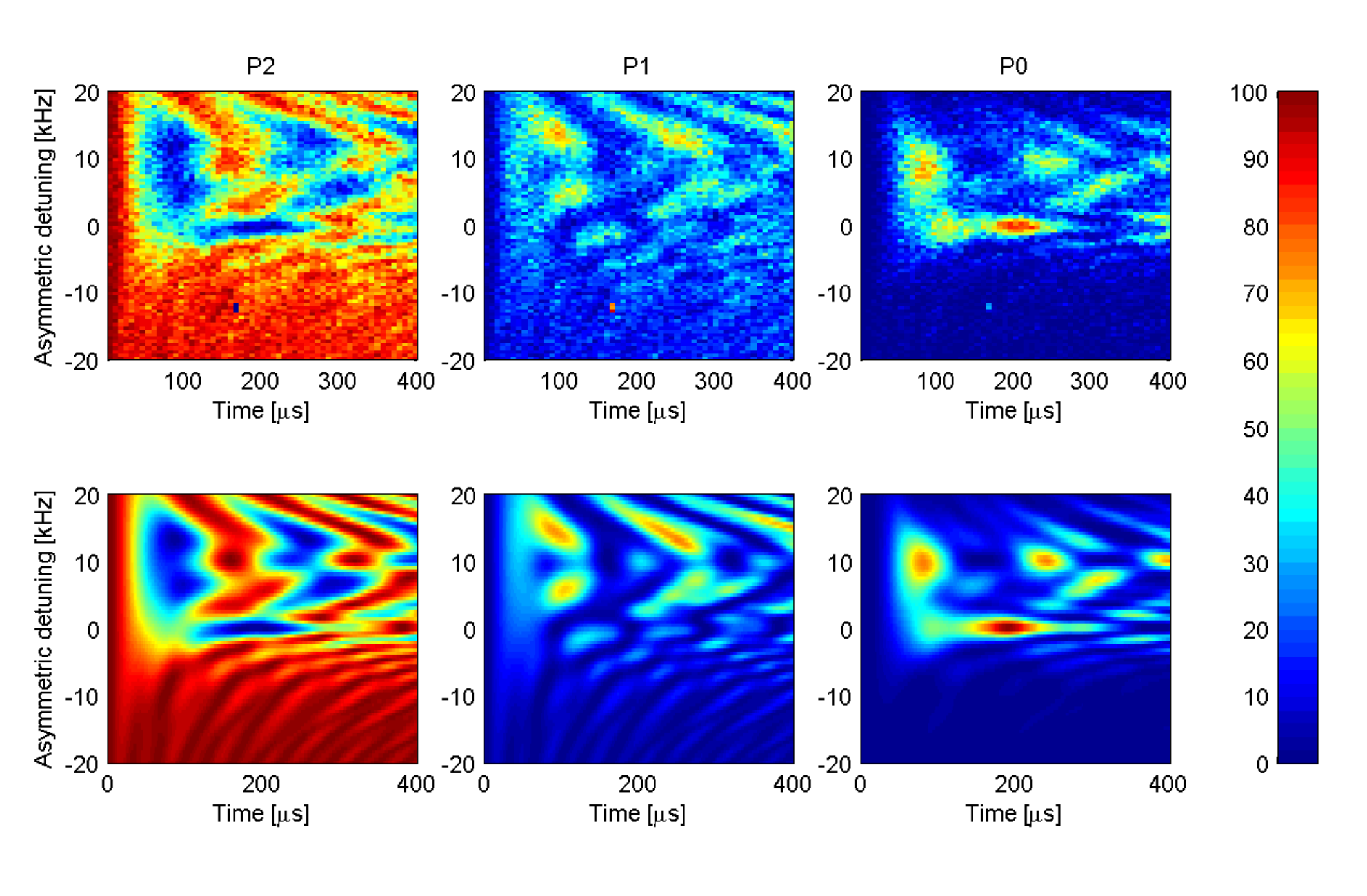}\\
  \caption{Populations time dynamic for the MS-interaction as function of the gate asymmetric detuning $\Delta$ with the optimal gate symmetric detuning $\delta/2\pi=10.5$ kHz. The calculated result shown by the lower panels was obtained by numerically  integrating the propagator of the Hamiltonian in Eq.(1).}
  \label{GateMapASymetric}
\end{figure}
\begin{figure}
  \center
  \includegraphics[width=15cm]{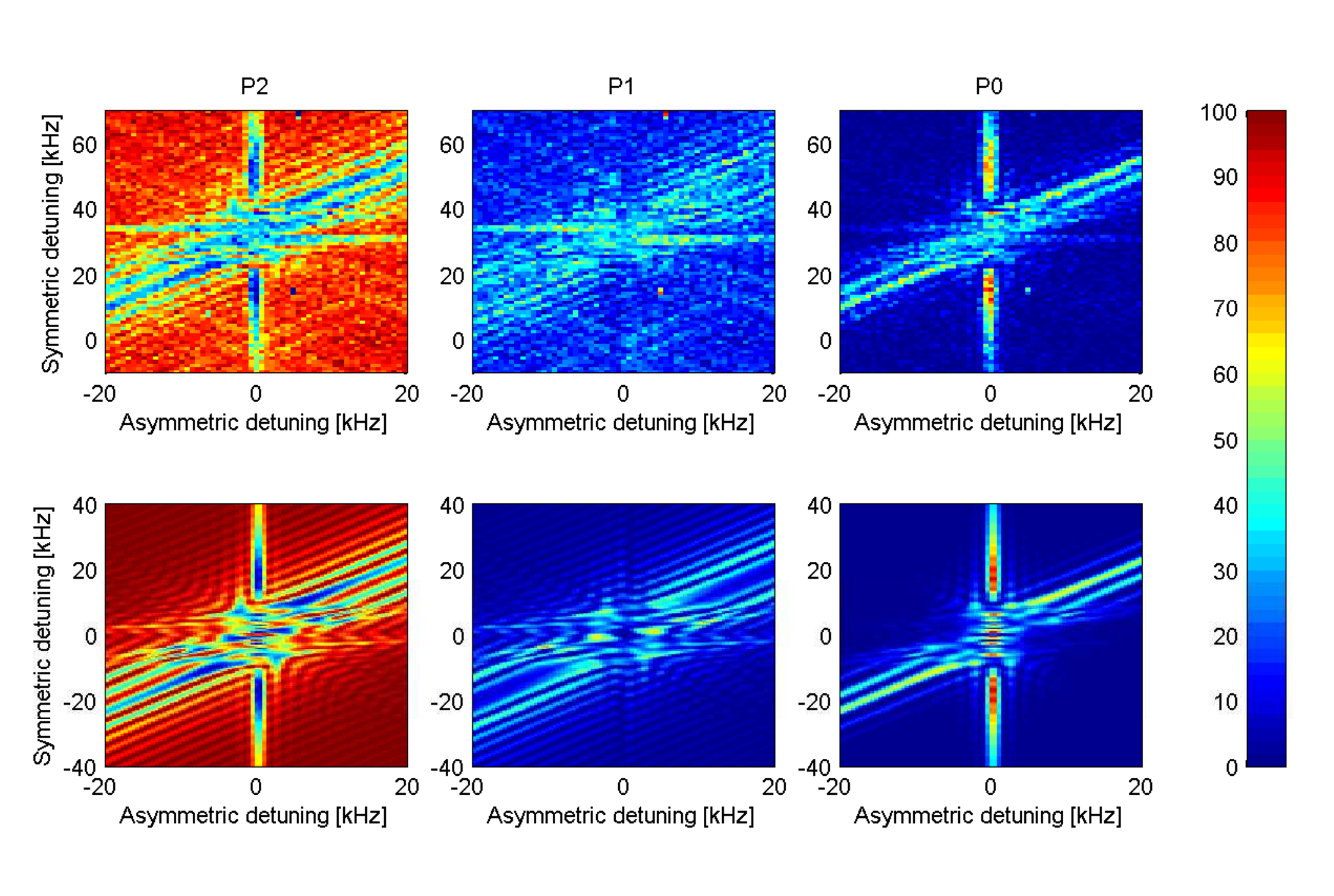}\\
  \caption{Populations for the MS-interaction as function of the gate symmetric detuning and asymmetric detuning for a fixed interaction time t=95 $\mu$S which is the gate-time for the appropriate detuning, $\delta/2\pi=10.5$ kHz and $\Delta=0$.}
  \label{GateMapUnionJack}
\end{figure}
Our experimental implementation of the MS-interaction is similar to that outlined in ref \cite{kirchmair2009deterministic}. A bichromatic beam was produced by single-passing an AOM(AOM-S) which was driven by two rf frequencies (RSB and BSB DDS sources). The beam then propagated to the ions through a polarization maintaining single mode fiber after which it was focused to a waist of about 20 $\mu$m. At the trap center each spectral component had up to 1 mW of power at 674 nm.  At each experiment the two ions were initialized to the $\ket{SS}$ state and their stretched mode which was used for the MS-interaction, was cooled to the ground state $(\bar{n}\approx 0.05)$. 

As a first step, we examined the ions state population for a wide interval of the interaction parameters: symmetric detuning $\delta$, asymmetric detuning $\Delta$ and interaction duration $t$. Here the center of mass vibrational mode with frequency of $\nu=0.98$ MHz was used. These scans provide a visual insight to the population dynamics and its dependence of the various parameters of the interaction. In addition, some experimental uncertainties in the gate parameters can substantially effect the gate fidelity. As an example, uncertainties in the intensities of each component in the bichromatic beam, the vibrational mode frequency and light shift from other levels affect the required gate time and detuning. As shown in \cite{kirchmair2009deterministic}, the experimental calibration of all these parameters using the gate itself is possible. A different approach, which is more straightforward, however time consuming, is to scan the relevant parameters: symmetric detuning $\delta$, asymmetric detuning $\Delta$ and time $T_{gate}$ and measure the two-ion response. Once the experimental maps are obtained the real parameters can be by extracted by comparing the measured maps to the calculated ones. 
Figures \ref{GateMapSymetric}-\ref{GateMapUnionJack} present the states population maps, both measured and calculated, for three 2D scans: symmetric detuning vs. time, asymmetric detuning vs. time and asymmetric vs. symmetric detuning.    

The practical value of such maps is best demonstrated in the time and symmetric detuning scan which is shown in Fig.\ref{GateMapSymetric}. The calculated populations as function of the gate time and symmetric detuning fully describe the measurement result up to a shift in the detuning of 35 kHz. This large shift is attributed to a bias in our  vibrational frequency measurement scheme. Our sideband frequency was measured via sideband spectroscopy using intense beams which generated light shift on the carrier transition frequency which was different in the MS gate beam configuration. Therefore, the measured sideband frequency included a significant light shift of 35 kHz.     

\subsection{Two-qubit entangling gate}
\begin{figure}
  \centering
  \includegraphics[width=15cm]{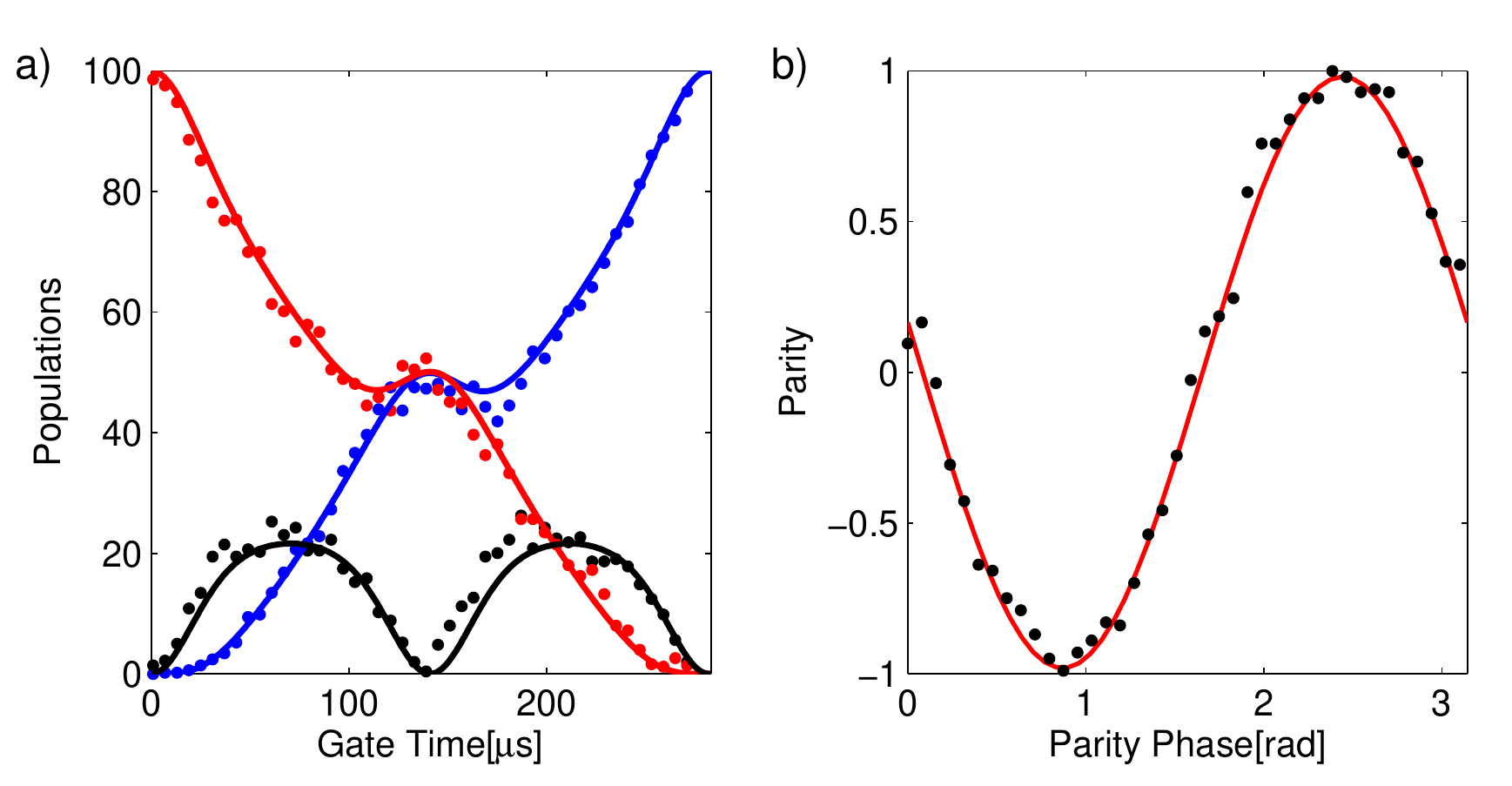}\\
  \caption{A M{\o}lmer-S{\o}rensen two-qubit gate after optimizing the gate parameters. (a) Populations time evolution as function of the interaction time. (b) Parity oscillation $P=P0+P2-P1$ obtained by scanning the phase of a $\pi/2$ pulse at the end of the gate at a gate-time $t_g$=130 $\mu S$. From these result we infer gate fidelity of F=0.985(10).}
  \label{GatePopAndParity}
\end{figure}

After a coarse setting of the gate parameters. Another set of quick one-dimensional scans were conducted in order to fine-tune the gate parameters and maximize the gate fidelity. First we scanned the gate symmetric detuning at the presumed gate time and set it to a value where the populations P0$\approx$P2 $\approx$50\%. Next, the gate time was scanned with the obtained gate detuning. Here the gate time was set to the point where P1 is minimal (ideally zero). The two scans were repeated several times  to make sure we converge to the correct values.         
Figure \ref{GatePopAndParity}a. presents the populations as a function of the interaction time. Here, we used the stretch vibrational mode . At a gate time of $T_{gate}=140~\mu$S the ions superposition consists of only the $\ket{SS}$ and $\ket{DD}$ states. To verify that the resulting state is indeed a Bell state, the off-diagonal elements of the density matrix must be measured as well. This was done by applying another $\pi/2$ pulse with a phase $\phi$ and measuring the state parity $P=P0+P2-P1$. The amplitude of parity oscillations, $A_p$, as function of $\phi$ is equal to $2|\rho_{SS,DD}|$. 
The gate fidelity for producing the Bell state $\ket{\Phi}$ is given by $F_\Phi=(P0+P2)/2+A_p$. Using the maximum likelihood method  to fit the data in Fig.\ref{GatePopAndParity}b, we conclude gate fidelity of F=0.985(10)\footnote{The likelihood of the measured data was calculated given a series of binomial distributions with mean values that follow a sinusoidal pattern with a given amplitude and phase. The amplitude and phase that yield the highest likelihood were used as an estimate for the gate fidelity.}.
While part ($\approx5\cdot10^{-3}$) of the gate error can be attributed to imperfection in the preparation and detection operations, the rest is assumed to arise from error in the gate itself. 
Although the exact source of error has not been fully identified, the nature of the error has been studied in \cite{navon2013quantum} where we have performed process tomography on five consecutive gates. The results were consistence with a collective depolarization channel. A possible source of error is off-resonance $\ket{S}\leftrightarrow \ket{D}$ incoherent transition as indicated by the measurement in the inset of Fig.\ref{MasterSpectrum}b.

\section{Summary}
We presented a universal gate-set for qubits encoded in the electronic states of $^{88}$Sr$^+$ ions. The gates were implemented  using a narrow linewidth diode laser. Diode lasers have a relatively wide frequency noise spectrum to begin with and therefore are challenging when trying to reach the narrow linewidths required. In particular, we have observed that unlike the case of optical atomic clocks, where noise faster than the linewidth of the laser plays a marginal role, here, residual fast phase noise in the servo-bumps that overlaps the motional sideband transition can generate significant errors to quantum gates.  
We show that one can overcome this problem by using a high-finesse cavity as a narrow optical filter. This moderates the servo bandwidth requirement and makes an ECDL a compact and affordable solution as a narrow-linewidh laser for quantum information processing.
We characterized the laser frequency uncertainty using the ions and found a linewidth below 100 Hz. In order to eliminate the contribution of the magnetic field noise from the result of a Ramsey experiment we introduced a new type of dynamic modulation sequence.  This method may find use in other applications such as atomic clocks that are not based on a magnetic insensitive transition \cite{NPLSrClock,NRCSrClock}.
We demonstrated high-fidelity collective single-qubit gates as well as individual qubit addressing in a two-qubit register by utilizing the inhomogeneous MM-sidebands. Although, this method for individual addressing  is limited for only two qubits and not applicable for scalable quantum information processing, we still find it extremely useful in many cases where only two qubits are of interest. For example, we have successfully used this method to measure the magnetic dipole interaction between the spins of two trapped ions \cite{kotler2014measurement}. 
We have implemented a M{\o}lmer-S{\o}rensen two-qubit gate with fidelity of F=0.985(10), thus realizing  a universal gate set for quantum computation with trapped ion qubits.
      
\section*{References}

\bibliography{UniversalGateBib}
\bibliographystyle{iopart-num}
\end{document}